\documentclass[twocolumn,showpacs,superscriptaddress,nobalancelastpage,amsmath,10pt,prl,a4paper]{revtex4} 

\usepackage{graphicx}

\DeclareMathOperator{\Tr}{Tr}

\begin{document}

\title{Calculation of the quantum discord in many-qubit systems}

\author{A.Y. Chernyavskiy}
\affiliation{Institute of Physics and Technology of the Russian Academy of Sciences}

\author{S.I. Doronin}
\email{s.i.doronin@gmail.com}
\affiliation{Institute of Problems of Chemical Physics of the Russian Academy of Sciences}

\author{E.B. Fel'dman}
\affiliation{Institute of Problems of Chemical Physics of the Russian Academy of Sciences}

\begin{abstract}
For the first time, we compute the quantum discord in bipartite systems containing up to nine qubits. 
An analytical expression is obtained for the discord in a bipartite system with three qubits. 
The dependence of the discord on the temperature and the structural parameter of the model is studied.
\end{abstract}


\maketitle

\section{Introduction}
The discord \cite{OZ, HV} is a measure of quantum correlations which are responsible for the effective work of quantum devices 
and give them significant advantages over their classical counterparts \cite{AZFY}. 
Calculations of the discord constitute a very laborious  optimization problem. As a result, the discord problem has been solved for two-qubit systems 
\cite{L,ARA,GA} and some simplest three-qubit systems \cite{RS} only. In the present work we study the quantum discord in bipartite many-qubit systems.

Nuclear magnetic resonance (NMR) is widely used for experimental realization of quantum algorithms in spin systems with 
a small number of qubits \cite{NC}. Untill recently entanglement was used as a sole measure of quantum correlations. Entanglement is small in liquid-state NMR experiments \cite{W}. 
So it seemed that the NMR methods were not promising for quantum information processing in many-qubit systems. A negative evaluation of  NMR methods 
has changed after it was understood that the quantum discord describes all quantum correlations whereas entanglement describes only a part of them. 
Discord can be significant in nuclear spin systems but is not well understood quantitatively yet.
Thus, methods of discord calculation in many-qubit systems became an important  problem now, in particular, in the NMR context.

\section{Model}
We consider a homogeneous chain  of nuclear spins ($S=1/2$) coupled by the dipole-dipole interactions (DDI) in a strong external magnetic field. 
We will call it subsystem $A$. Subsystem $A$ interacts with an impurity spin $S$ (subsystem $B$, Fig.1).

\begin{figure}[ht]
\includegraphics[width=0.95\columnwidth]{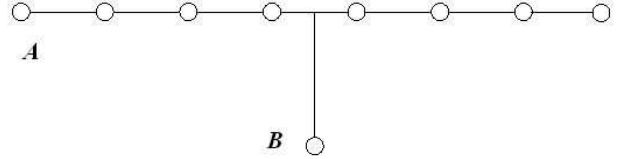}
\caption{\label{fig:F1} The many-spin system for the investigation of the discord. 
Subsystem $A$ is a linear chain of spins coupled by the DDI. The impurity spin, connected with the chain spins 
by $zz$- interactions, represents subsystem $B$.}
\end{figure}

In the initial moment of time the system is in the thermodynamic equilibrium state with the density matrix
\begin{equation}
\rho(0) = \frac{1}{Z}e^{\beta(\omega_A I_z + \omega_B S_z)},\label{eq:1}
\end{equation}
where $z$ points in the direction of the magnetic field which is perpendicular to the plane of Fig.1; $I_z=\sum_iI_{zi}$, $I_{zi}$ and $S_z$
are the $z$- projections of the $i$-th chain spin and the impurity spin; $\omega_A$ and $\omega_B$ are the Larmour frequencies; 
$\beta$ is proportional to the inverse temperature of the system and $Z$ is the partition function.
Applying resonance
 (for spins $I$ and $S$) $90^\circ$- radiofrequency pulses of the magnetic field around the axis $y$ in the rotating reference frame \cite{G}
 we create conditions for the emergence of quantum correlations in the course of the evolution of the spin system. 
Formally such initiating pulses lead to the replacement of the operators $I_{iz}, S_z$ with $I_{ix}$ and $S_x$ 
in the expression for the density matrix (\ref{eq:1}).
The subsequent evolution of the system is described by the Hamiltonian $\tilde{H}=H_{dz}+H_{zz}$. 
Here the Hamiltonian $H_{dz}$ describes the DDI in the spin chain (subsystem $A$) and is determined by the secular part of the DDI as \cite{G}
\begin{equation}
H_{dz}=\sum_{i<j} d_{ij}\left(3I_{iz}I_{jz} - \vec{I_i}\vec{I_j}\right),\label{eq:2}
\end{equation}
where $i,j$ are the chain spins; $d_{ij}$ is the DDI coupling constants of spins $i$ and $j$; $\vec{I_i}\vec{I_j}=I_{ix}I_{jx}+I_{iy}I_{jy}+I_{iz}I_{jz}$. 
In turn the Hamiltonian $H_{zz}$ characterizes the interaction between subsystems $A$ and  $B$. It equals
\begin{equation}
H_{zz}=\sum_{i} g_{i}I_{iz}S_{z},\label{eq:3}
\end{equation}
where $g_{i}$ is the coupling constant of the chain spin $i$ with the inpurity spin which has a different gyromagnetic ratio from the chain spins.

One can find one-dimensional chains of nuclear spins coupled with the DDI, for example, in calcium hydroxyapatite $Ca_5(OH)(PO_4)_3$ 
(chains of hydroxyl protons) and in calcium fluorapatite $Ca_5F(PO_4)_3$ (chains of fluorine nuclei) \cite{CY}. 
In those substances the distances between neighboring parallel chains is three times greater than the distance between nearest neighbors in the chains \cite{CY}. This circumstance is a motivation for the model of the isolated spin chain \cite{FL} coupled with the spins of impurity nuclei.

The evolution of the density matrix in the described model can be investigated by NMR methods \cite{G}. It yields ways for discord measurements in solids.
In particular, NMR quantum state tomography \cite{OBS} can be used. The suggested model can be considered as a theoretical bases for the interpretation of the results of these experiments.

\section{The discord in the tree-qubit system}

First we consider a case where subsystem $A$ contains only two spins. 
It is suitable to rewrite the initial density matrix after applying $90^\circ$- radiofrequency pulses as 

\begin{equation}
\begin{array}{l}  
            \rho(0) = \frac{1}{Z}\Big[\cosh^2{(\frac{\beta\omega_A}{2})}+\sinh{(\beta\omega_A)}I_x\\
+4\sinh^2{(\frac{\beta\omega_A}{2})}I_{1x}I_{2x}\Big]\left[\cosh{\frac{\beta\omega_B}{2}}+2\sinh{(\frac{\beta\omega_B}{2})}S_x\right],\label{eq:4}
      \end{array}
\end{equation}
$$Z=8\cosh^2{\frac{\beta\omega_A}{2}\cosh{\frac{\beta\omega_B}{2}}}.$$

One can find from the symmetry of the system (see Fig.1) that the Hamiltonians $H_{dz}$ and $H_{zz}$ commute. 
Thus, subsystem $A$ is only subject to dipolar evolution. Since a unitary local transformation does not change the discord, 
we can ignore dipolar evolution of subsystem $A$. Assuming that $g_1=g_2=g$ one can find

\begin{equation}
\begin{array}{l} 
e^{-iH_{zz}t}S_xe^{iH_{zz}t}=\left[\cos^2{\frac{gt}{2}}-4\sin^2{\frac{gt}{2}}I_{1z}I_{2z}\right]S_x\\
+(I_{1z}+I_{2z})S_y\sin{(gt)},\label{eq:5}
      \end{array}
\end{equation}
where $t$ is the evolution time.

In the course of the evolution determined by the Hamiltonian $H_{zz}$ the density matrix of the system $\rho(t)$ gets the form

\begin{equation}
\begin{array}{l} 
\rho(t)=e^{-iH_{zz}t}\rho(0)e^{iH_{zz}t}=\frac{1}{Z}\Big\{\cosh{(\frac{\beta\omega_B}{2})}A\\
+S_z\cosh{(\frac{\beta\omega_B}{2})}B+S_x\Big[2\sinh{(\frac{\beta\omega_B}{2})}A\\
\times\Big(\cos^2{\frac{gt}{2}}-4\sin^2(\frac{gt}{2})I_{1z}I_{2z}\Big)\\
-i\sinh{(\frac{\beta\omega_B}{2})}\sin(gt)BI_z\Big]+S_y\Big[2\sinh{(\frac{\beta\omega_B}{2})}\sin(gt)AI_z\\
+i\sinh{(\frac{\beta\omega_B}{2})}B\Big(\cos^2{(\frac{gt}{2})}-
4\sin^2(\frac{gt}{2})I_{1z}I_{2z}\Big)\Big]\Big\},\label{eq:6}
      \end{array}
\end{equation}
where
\begin{equation}
\begin{array}{l} 
A=\cosh^2{(\frac{\beta\omega_A}{2})+\sinh(\beta\omega_A)\cos{(\frac{gt}{2})I_x}}\\
+\sinh^2{(\frac{\beta\omega_A}{2})}\Big[2\cos^2{(\frac{gt}{2})}I_x^2+2\sin^2{(\frac{gt}{2})I_y^2-1\Big]},\label{eq:7}
      \end{array}
\end{equation}
\begin{equation}
\begin{array}{l} 
B=2\sinh(\beta\omega_A)\sin(\frac{gt}{2})I_y-i\sinh^2{(\frac{\beta\omega_A}{2})}\\
\qquad \times\sin(gt)\Big[(I^+)^2-(I^-)^2\Big],\label{eq:8}
      \end{array}
\end{equation}
and $I^+,I^-$ are the raising and lowering spin angular momentum operators. 

In order to calculate the discord we need the reduced density matrices of subsysthems $A$ and $B$:
\begin{equation}
\begin{array}{l} 
\rho_A(t)=\Tr_B\rho(t)=\frac{2}{Z}\cosh(\frac{\beta\omega_B}{2})A,\label{eq:9}
      \end{array}
\end{equation}
\begin{equation}
\begin{array}{l} 
\rho_B(t)=\Tr_A\rho(t)=\frac{1}{2}+\tanh(\frac{\beta\omega_B}{2})\cos^2(\frac{gt}{2})S_x,\label{eq:10}
      \end{array}
\end{equation}
and also the von Nemann entropy of the whole bipartite system $S(\rho)$ and the entropies $S(\rho_A)$ and $S(\rho_B)$ of its subsystems:
\begin{equation}
\begin{array}{l} 
S(\rho)=3+\frac{2\ln(\cosh\frac{\beta\omega_A}{2})}{\ln2}+\frac{\ln(\cosh\frac{\beta\omega_B}{2})}{\ln2}\\
\qquad -\frac{\beta\omega_A\tanh\frac{\beta\omega_A}{2}}{\ln2}-\frac{\beta\omega_B\tanh\frac{\beta\omega_B}{2}}{2\ln2}, \label{eq:11}
      \end{array}
\end{equation}
\begin{equation}
\begin{array}{l} 
S[\rho_A(0)]=2+\frac{2\ln(\cosh\frac{\beta\omega_A}{2})}{\ln2}-\frac{\beta\omega_A\tanh\frac{\beta\omega_A}{2}}{\ln2}, \label{eq:12}
      \end{array}
\end{equation}
\begin{equation}
\begin{array}{l} 
S(\rho_B)=1-\frac{1}{2\ln2}\Big[(1+\tanh\frac{\beta\omega_B}{2}\cos^2\frac{gt}{2})\\
\times\ln(1+\tanh\frac{\beta\omega_B}{2}\cos^2\frac{gt}{2})+(1-\tanh\frac{\beta\omega_B}{2}\cos^2\frac{gt}{2})\\
\times\ln(1-\tanh\frac{\beta\omega_B}{2}\cos^2\frac{gt}{2})\Big]. \label{eq:13}
      \end{array}
\end{equation}

We use the entropy $S[\rho_A(0)]$ at the initial moment of time because it is sufficient for the calculation of the discord (see below). 
According to the standard approach \cite{L,ARA}, one performs a total set of projective measurements over one-qubit subsystem $B$ 
$$
\{B_k=V\Pi_kV^\dagger, \quad V\in U(2), \quad \Pi_k=\left|k\right\rangle\left\langle k\right|, \quad k=0, 1\}.
$$
After the measurements, one can find that the whole system is then described by the ensemble of the states $\{p_k,\rho_k\} (k=0, 1)$:
\begin{equation}
\begin{array}{l}
p_0=\frac{1}{2}+\frac{z_1}{2}\tanh\frac{\beta\omega_B}{2}\cos^2\frac{gt}{2},\\
p_1=\frac{1}{2}-\frac{z_1}{2}\tanh\frac{\beta\omega_B}{2}\cos^2\frac{gt}{2}, \label{eq:14}
     \end{array}
\end{equation}
\begin{equation}
\begin{array}{l} 
\rho_k=\frac{2}{Z(1\pm z_1\tanh\frac{\beta\omega_B}{2}\cos^2\frac{gt}{2})}\Big\{\cosh{(\frac{\beta\omega_B}{2})}A\pm \frac{z_3}{2}\cosh{(\frac{\beta\omega_B}{2})}B\\
\pm \frac{z_1}{2} \Big[2\sinh{(\frac{\beta\omega_B}{2})}A\Big(\cos^2{\frac{gt}{2}}-4\sin^2(\frac{gt}{2})I_{1z}I_{2z}\Big)\\
-i\sinh{(\frac{\beta\omega_B}{2})}\sin(gt)BI_z\Big]\pm \frac{z_2}{2}\Big[2\sinh{(\frac{\beta\omega_B}{2})}\sin(gt)AI_z\\
+i\sinh{(\frac{\beta\omega_B}{2})}B\Big(\cos^2{(\frac{gt}{2})}-4\sin^2(\frac{gt}{2})I_{1z}I_{2z}\Big)\Big]\Big\}, \,\,(k=0, 1), \label{eq:15}
      \end{array}
\end{equation}
where the parameters $z_1, z_2, z_3 \, (z_1^2+z_2^2+z_3^2=1)$ characterize different unitary transformations used at the minimization of the entropy after measurements over subsystem $B$. 
These parameters should be found from the minimization of the entropy of the system after the measurements, $S(\rho \,|\{B_k\})$, 
which is determined by \cite{L,ARA}:
\begin{equation}
S(\rho \,|\{B_k\})=p_0S(\rho_0)+p_1S(\rho_1). \label{eq:16}
\end{equation}

\begin{figure}[t]
\includegraphics[width=0.65\columnwidth]{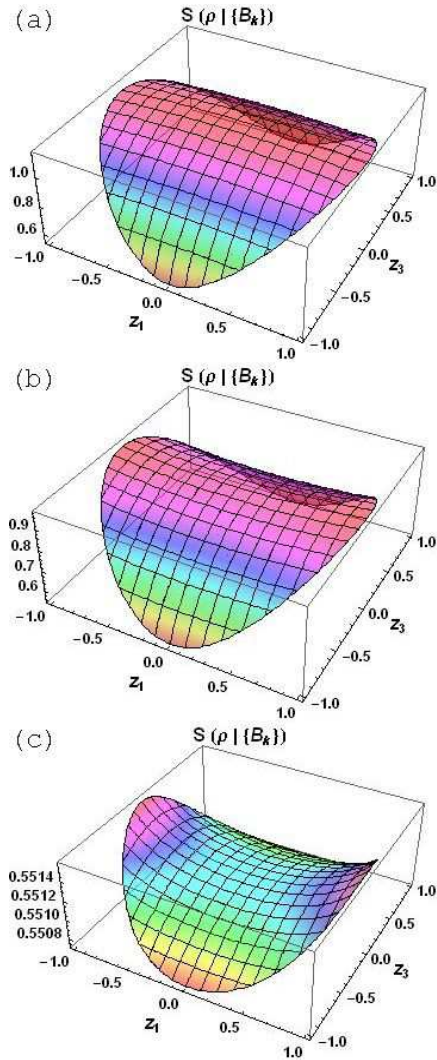}
\caption{\label{fig:F2} The dependence of the conditional entropy $S(\rho \,|\{B_k\})$ in the three-qubit system on the parameters $z_1, z_3$, used at the minimization \cite{L,ARA} at different moments of the evolution time $t$: (a) $t=1$ ms, (b) $t=1.5$ ms, (c) $t=2$ ms. The coupling constant of the $zz$- interactions $g=3120.79 s^{-1}$.}
\end{figure}

We performed a numerical calculation of the conditional entropy $S(\rho \,|\{B_k\})$ at different moments as a function of the 
parameters $z_1$ and $z_3$ (Fig.2). The minimal value of the conditional entropy $S(\rho \,|\{B_k\})$ is achieved at $z_1=z_2=0, |z_3|=1$ for every moment of time  (Fig.2).

Performing the transformation
\begin{equation}
\tilde{\rho}_k=e^{i(-1)^k\frac{gt}{2}I_z}\,\rho_k\,e^{-i(-1)^k\frac{gt}{2}I_z}, \quad k=0,1 \label{eq:17}
\end{equation}
at $z_1=z_2=0, |z_3|=1$ in (\ref{eq:15}) one can find that the matrices $\tilde{\rho}_k\,(k=0,1)$ do not depend on time. 
It means that the eigenvalues of the matrices $\rho_k\ (k=0,1)$ which determine the conditional entropy $S(\rho \,|\{B_k\})$
 do not depend on time too. Further we consider the matrices $\rho_k\,(k=0,1)$ setting  $t=0$ in Eq. (\ref{eq:15}).
 Since one can obtain from (\ref{eq:8}) that $B=0$ at $t=0$, it is possible to conclude that $\rho_0=\rho_1=\rho_A(0)$ (see Eq.(\ref{eq:9})). 
Now we can obtain from Eqs. (\ref{eq:14}), (\ref{eq:16}) that $S(\rho \,|\{B_k\})=S(\rho_A(0))$. 

Total quantum correlations are determined by mutual information \cite{V}
\begin{equation}
I(\rho)=S(\rho_A)+S(\rho_B)-S(\rho). \label{eq:18}
\end{equation}
After performing the total set of measurements over subsystem $B$, 
one can write the expression for mutual information which differs from (\ref{eq:18}) in the quantum case
\begin{equation}
\tilde{I}(\rho)=S(\rho_A)-\underset{\{B_k\}}{\min}{S(\rho \,|\{B_k\})}. \label{eq:19}
\end{equation}
The difference $D$ (discord) determines quantum correlations in the system:
\begin{equation}
\begin{array}{l} 
D=I(\rho)-\tilde{I}(\rho)=-\frac{1}{2\ln2}\Big[(1+\tanh\frac{\beta\omega_B}{2}\cos^2\frac{gt}{2})\\
\times\ln(1+\tanh\frac{\beta\omega_B}{2}\cos^2\frac{gt}{2})+(1-\tanh\frac{\beta\omega_B}{2}\cos^2\frac{gt}{2})\\
\times\ln(1-\tanh\frac{\beta\omega_B}{2}\cos^2\frac{gt}{2})\Big]-\frac{\ln(\cosh{\frac{\beta\omega_B}{2}})}{\ln2} + \frac{\beta\omega_B}{2\ln2}\tanh\frac{\beta\omega_B}{2}. \label{eq:20}
   \end{array}
\end{equation}

\begin{figure}[t]
\includegraphics[width=0.95\columnwidth]{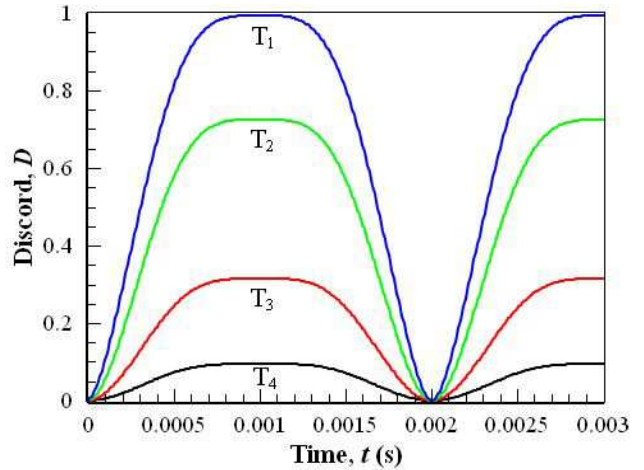}
\caption{\label{fig:F3} The dependence of the discord $D$ on the evolution time at different temperatures $(T_1=0.0001\,K, T_2=0.0005 \,K, T_3=0.001 \,K, T_4=0.002 \,K)$. The dimensionless parameter $\beta\omega_B=0.015/T$ ($T$ is the temperature), $g=3120.79 s^{-1}$.}
\end{figure}

The dependence of the discord $D$ of the three-qubit system on the temperature is shown in Fig.3. 
 Quantum correlations decay and the discord decreases as the temperature increases. The discord equals zero at all temperatures, if $gt=2\pi$ (see Fig.3).

\section{The discord in many-qubit systems}
Our optimization procedure for the discord in five-, seven- and nine-qubit systems uses an algoritm with random mutations. 
\begin{figure}[t]
\includegraphics[width=0.65\columnwidth]{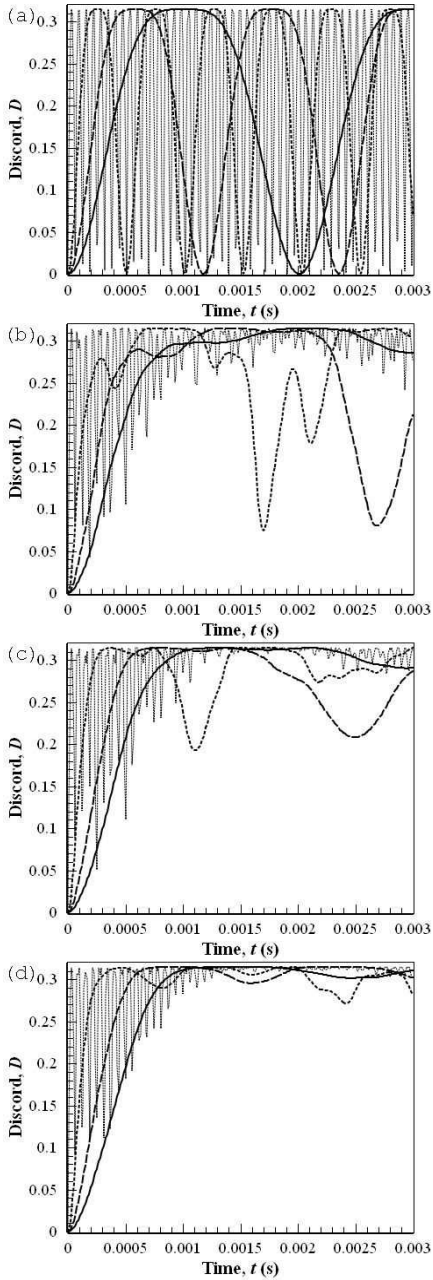}
\caption{\label{fig:F4} The discord evolution in bipartite systems at different distances $b$ of the impurity spin from the chain. The distance $a$ between the nearest chain spins is fixed. Thin  dots correspond to $b^2/a^2=0$, bold  dots correspond to $b^2/a^2=0.75$, dash lines correspond to $b^2/a^2=1.5$, solid lines correspond to $b^2/a^2=2.25$: (a) The system consisting of $2+1$ qubits; (b) the system consisting of $4+1$ qubits; (c) the system consisting of $6+1$ qubits; (d) the system consisting of $8+1$ qubits. }
\end{figure}
It is related to a genetic algoritm \cite{Ch}. In all cases we have found that the minimal value of the conditional entropy \cite{L,ARA} does not depend on time.
For a given number of the qubits the discord 
evolution is investigated at different distances of the impurity spin from the chain.
The evolution of the discord in bipartite systems consisting of $2+1, 4+1, 6+1, 8+1$ qubits is given in Fig.4 and exhibits oscillatory behavior.  

For a system with $2+1$ qubits the oscillations (see Fig.4a) are determined by the frequency $g$.  
This frequency decreases when the impurity spin moves from the chain. Correspondingly, the oscillation period increases (Fig.4a). 
In systems with greater numbers of qubits the oscillations are determined by a set of frequencies $g_i$. As a consequence, the oscillations of the type shown in Fig. 4a are smeared.
The DDI of the chain spins also affect  the discord evolution (see Fig.4b,c,d). All our calculations are performed at the temperature $T=0.001 K$ where we obtain large values of the discord. It means that there are significant quantum correlations in the considered model at low temperatures.
\section{Conclusion}
The model of the chain of nuclear spins coupled by the DDI and interacting with the impurity spin allows us to investigate the discord evolution 
(the evolution of quantum correlations) in many-qubit systems. We found an analytical expression for the discord in a bipartite system consisting of three qubits. 
The algorithm with random mutations is effective at the investigation of the discord and its evolution in many-qubit systems.

The authors thank  A.I. Zenchuk and M.A. Yurishchev for useful discussions. 
The work is supported by the Russian Foundation for Basic Research (Grants No. 13-03-00017 and No. 12-01-31274) and the Program of the Presidium of RAS No. 8 ``Developments of methods of obtaining chemical compounds and creation of new materials''.

\end{document}